\documentstyle[aps,preprint]{revtex}
\begin{document}
\bibliographystyle{unsrt}

\title{Incremental expansions for the  ground state energy
of the two-dimensional
Hubbard model} 

\author{ Jiri Malek$^*$ and Sergej Flach}

\address{Max-Planck-Institute for Physics of Complex Systems, Noethnitzer Str. 38, 
D-01187 Dresden, Germany 
}

\author{ Konstantin Kladko}
\address{Center for Nonlinear Studies, Los Alamos National Laboratory,
Los Alamos, NM 87545, USA}
\date{\today}
\maketitle
$^*$ on leave of absence from: Institute of Physics of AVCR,
18221 Prague 8, Na Slovance 2, Czech Republic
\begin{abstract}
A generalization of Faddeev's approach of the 3-body problem
to the many-body problem 
leads to the method of increments.
This method was recently applied to account for the ground state
properties of Hubbard-Peierls chains (JETP Letters
67 (1998) 1052).
Here we generalize this approach to two-dimensional
square lattices and explicitely treat the incremental
expansion up to third order. 
Comparing our numerical results with 
various other approaches (Monte Carlo, cumulant approaches)
we show that incremental expansions are very efficient because
good accuracy with those approaches 
is achieved treating lattice segments composed
of 8 sites only.
\end{abstract}
\newpage


The understanding of properties of strongly interacting fermions
has been an intense topic of research for the past decade, in part
due to the interest in the properties of high temperature superconductors.
As exact solutions are known only for selected integrable models,
numerical methods gained importance to provide benchmarks for
analytical approaches which necessarily use approximations of all
kinds. 

Incremental expansions have been used in quantum chemistry
to account for properties of molecules and solids \cite{pf95}. 
In a recent work
\cite{kf98} these methods were combined with cumulant expansions to
provide a solid footing for numerical implementations. One result
was, that incremental expansions can be interpreted as
approximative ways to solve Faddeev-like equations for the N-body
problem. 

To explain this in more detail, let us consider a Hamiltonian 
\begin{equation}
H=H_0+\sum_n H_n \;\;.
\label{1-1}
\end{equation}
Assume that we can find the eigenenergies and eigenvectors
of $H_0$ and of $H_0+H_n$ for any $n$. Suppose further that
$S_n$ is the scattering operator associated with $H_0+H_n$. Then
Faddeev-like equations read as \cite{kf98}
\begin{equation}
T_n=S_n(1+\sum_{m\neq n}T_j)\;\;,\;\;S=\sum_n T_n \;\;,
\label{1-2}
\end{equation}
where $S$ is the scattering operator of the full Hamiltonian $H$.
Assuming the $S_n$ to be small, one obtains in lowest order $T_n=S_n$
and $S=\sum_n S_n$.
Having the $S$ operator one can calculate e.g. the ground state energy
of $H$ (for details see \cite{kf98}). 

The ground state energy $E$ in zero order of the incremental
expansion equals to the ground
state energy $E^{(0)}$ of $H_0$. In first order we take into account
the effect coming from adding one single $H_n$. Denote the ground state
energy of $H_0 + H_n$ by $E_n$. Then the first order increment to
the ground state energy is given by $I_n^{(1)}=E_n-E^{(0)}$. This
increment measures the change of adding $H_n$ to the ground state
energy. Summing up all increments of first order (for different $n$)
we obtain the ground state energy $E^{(1)}$ to first order
\begin{equation}
E^{(1)} = E^{(0)} + \sum_n I_n^{(1)} \;\;. 
\label{1-3}
\end{equation}
Let us now calculate the second order increments, which are
taking into account the simultaneous effect of adding two terms $H_n$ and
$H_m$. Here one has to subtract the corrections coming from adding
both terms separately. Denote the ground state energy of $(H_0+ H_n + H_m)$
with $E_{nm}$. Then the second order increment is given by
$I_{nm}^{(2)} = E_{nm} - I_n^{(1)} - I_m^{(1)} - E^{(0)}$. The ground
state energy to second order $E^{(2)}$ is then given by
(see Chapter 5.2.3 in \cite{pf95})
\begin{equation}
E^{(2)} = E^{(0)} + \sum_n I_n^{(1)} + \sum_{n<m} I_{nm}^{(2)} \;\;.
\label{1-4}
\end{equation}
This procedure can be easily continued to higher orders, and
the ground state energy will be a sum over increments of all orders. 
It requires 
the exact calculation of $E_{nm}$ when going to second order (
respectively $E_{nml}$ when going to third order etc).
Clearly this procedure does not imply certain topological structures
induced by $\sum_n H_n$, so that we are not restricted to certain
space dimensions if considering spatially extended systems.
Note that there are many ways
to split a full Hamiltonian $H$ into a trivially solvable part $H_0$
and the terms $H_n$. Each of these ways would generate
its own incremental expansions. Finally these expansions are not
restricted to the calculation of the ground state energy only, but can
be also applied to other ground state properties
and to excited states \cite{kf98}.

This method was successfully applied to one-dimensional lattices
\cite{mkf98}. There the special topology of a 1d system
lead to the successive cancellation of lower order increments
when proceeding to higher orders. 
Here we use this method to account for the ground
state energy of the two-dimensional Hubbard model on
a square lattice at half filling.
The Hamiltonian in dimensionless units is given by
\begin{equation}
\label{2-1}
H_{el} = -\sum_{i,j,\sigma} (c_{i,j,\sigma}^{\dagger} c_{i+1,j,\sigma}
+ c_{i,j,\sigma}^{\dagger} c_{i,j+1,\sigma} +h.c.) +
U\sum_{i,j}  n_{i,j,\uparrow} n_{i,j,\downarrow}\;\;.
\end{equation}
Here $i$ and $j$ are integers (which denote the $x$ and $y$ coordinates
of the lattice points).
To proceed we have to define a splitting of  $H_{el}$ into $H_0$
and $\sum_n H_n$. Here we consider $H_0$ as given by $H_{el}$ when
all vertical bonds and each second horizontal bond are missing
(see also Fig.1):
\begin{equation}
H_0 = -\sum_{i,j,\sigma}  \left(
c_{2i+j,j,\sigma}^{\dagger} c_{2i+j+1,j,\sigma} 
+ h. c.\right) +
U\sum_{i,j}  n_{i,j,\uparrow} n_{i,j,\downarrow}\;\;.
\label{2-2}
\end{equation}
Thus $H_0$ is a set of horizontally alligned noninteracting
dimers. 
Note that $H_0$ already contains all correlation
terms of (\ref{2-1}). The terms $H_n$ are then the missing bonds
i.e. all vertical bonds 
$\sum_{i,j}(c_{i,j,\sigma}^{\dagger} c_{i,j+1,\sigma} +h.c.)$
and the missing horizontal bonds 
$\sum_{i,j}(c_{2i+j-1,j,\sigma}^{\dagger} c_{2i+j,j,\sigma}
+ h. c.)$.

For sake of concreteness let us assume that our initial 
model Hamiltonian (\ref{2-1}) has an even and finite number of $2N$ sites, and 
periodic boundary conditions. In zero order the ground state energy $E$
of the whole system and the energy per site $\epsilon$ are given by
\begin{equation}
E^{(0)}=N E(C_0) \;\;,\;\;\epsilon^{(0)}=\frac{1}{2} E(C_0) \;\;,
\label{2-4}
\end{equation}
where $E(C_0)$ is the ground state energy of the 0th configuration
$C_0$
which per definition is a dimer with two electrons (Fig.2).

The first order increment is given by adding one of the missing bonds.
The ground state energy of this case is given by sum over
the ground state energies of $N-2$ dimers ($C_0$) and
of an open segment of two coupled dimers $C_1$
 (see Fig.2):
\begin{equation}
I{(C_1)} =E(C_1) + (N-2)E(C_0) - N E(C_0) = E(C_1) - 2 E(C_0)\;\;.
\label{2-5}
\end{equation}
We will encode all configurations of linked dimers to be considered by $C_n$.
Note that all increments are independent of the given
position of the returned bond because
we assumed periodic boundary conditions. Next we need to account for 
the weight factor $w(C_1)$ of $I{(C_1)}$, 
i.e. the number of increments
per dimer of $H_0$ having the same energy. 
It is easy to see that $w(C_1)=3$ (all possible
realizations are shown in Fig.2). 
The ground state energy in first order is then
\begin{equation}
E^{(1)} = E^{(0)} + 3NI{(C_1)} = N (3E(C_1) - 5E(C_0))\;\;,\;\;
\epsilon^{(1)} = \frac{1}{2}(3E(C_1) - 5E(C_0))\;\;.
\label{2-6}
\end{equation}
Already at this stage, though the considered configurations
are equivalent to those of a 1d chain at the same order of
incremental expansions \cite{mkf98} we find a difference
in the energy per site due to the increased number of nearest neighbours
of the 2d lattice as compared to the 1d case.

In second order we have to add two of the missing bonds.
Nonzero contributions come from cases when the two returned
bonds are linked \cite{kf98}. Then we have two nonzero
configurations $C_2$ (open chain with six sites) and $C_3$
(which is already incorporating topological effects of the
2d system) in this order (see Fig.3). Their weight factors are
$w(C_2)=9$ and $w(C_3)=6$. The corresponding increments are
\begin{eqnarray}
I(C_2) = E(C_2) -3E(C_0) - 2I(C_1) = E(C_2) - 2E(C_1) +
E(C_0) \;\;, \label{2-7} \\
I(C_3) = E(C_3) - 3E(C_0) - 2I(C_1) = E(C_3) - 2E(C_1) + E(C_0)
\;\;. \label{2-8}
\end{eqnarray}
The ground state energy can be 
evaluated according to (\ref{1-4}):
\begin{equation}
\epsilon^{(2)} = \frac{1}{2}E(C_0) + \frac{3}{2} I(C_1) 
+ \frac{9}{2} I(C_2) + 6I(C_3) =
5E(C_0) - 13.5 E(C_1) + 4.5E(C_2) + 3 E(C_3) \;\;. \label{2-9}
\end{equation}
 Since the topology of the 
configurations starts to be different from those appearing in
a 1d system \cite{mkf98} no trivial cancellation of lower 
order increments takes place anymore. 

In third order we add three missing bonds (again only configurations
when all three returned bonds are linked do contribute). We obtain
six different configurations $C_4,C_5,C_6,C_7,C_8,C_9$ with corresponding
weight factors $2,27,2,10,18,32$. They are shown in Figs.3,4.
The corresponding increments are
\begin{eqnarray}
I(C_4) = E(C_4) - I(C_2) - 2 I(C_3) - 3 I(C_1) - 3 E(C_0)\;\;,
\label{2-10} \\
I(C_5) = E(C_5) - 2 I(C_2) - 3 I(C_1) - 4 E(C_0) \;\;, \label{2-11} \\
I(C_6) = E(C_6) - 3 I(C_3) - 3 I(C_1) - 4 E(C_0) \;\;, \label{2-12} \\
I(C_7) = E(C_7) - 2 I(C_3) - 3 I(C_1) - 4 E(C_0)\;\;, \label{2-13} \\
I(C_8) = E(C_8) - I(C_3) - 2 I(C_2) - 3 I(C_1) - 4 E(C_0)\;\;
\label{2-14} \\
I(C_9) = E(C_9) - I(C_2) - I(C_3) - 3 I(C_1) - 4 E(C_0)\;\;.
\label{2-15}
\end{eqnarray}
It is too tedious to explicitely write down the formula for the
ground state energy. In the following we will present the results
of numerical calculations.

We use a Lancosz algorithm to compute the ground state energies
of our considered configurations.
In Fig.5 we show the dependence of the ground state energy per site
$\epsilon$ on the interaction parameter $U$ 
in third order of the incremental expansion.
In the inset of Fig.5 we show the $U$-dependence of the different
incremental contributions. We find that the contributions coming
from 2nd and 3rd  order are small compared to the 0th and 1st order. 
For free electrons $U=0$ we compare $\epsilon$
for 0th,1st,2nd and 3d orders - $-1; -1.708204; -1.768700;
-1.6335775$
with the exact number $\epsilon(U=0) = -1.621139$ (see also Table 1 and
Fig.5).
This gives a relative error of only 0.8\% !
For $U=1,2,4$ 
we compare $\epsilon$ in 3d order in Table 1 with 
quantum Monte Carlo (QMC) calculations of Moreo et al \cite{msswb90},
where lattices with sizes up to $16\times 16$ were used and
extrapolations were carried out. The relative difference (Table 1) is less
than 2\%. Note that the slight increase in the relative error
with increasing $U$
at least partially has to be attributed to the circumstance
that the QMC calculations become less exact with increasing
$U$. 
Also in Fig.5 the  results of Polatsek and Becker on  
related projection operator techniques
using cumulants
are shown \cite{pb96}. 
These calculations take spin flips up to second
order and charge fluctuations up to second order (in terms
of our notations) into account. The agreement is very good in a
broad range of $U$ values. The exception is the limit of small $U$,
where the projection technique becomes less accurate.  
Finally for large $U$ the Hubbard model
transforms into the Heisenberg model. Doing the same calculations
for the Heisenberg model with $J=1$  we compare 
$\epsilon$ for 0th,1st,2nd and 3d orders - $-0.375; -0.5490375;
-0.578994; -0.6695330$ with the
results of QMC calculations of the 
ground state energy  of the Heisenberg model  $\epsilon
\approx -0.669$ by Runge \cite{kjr92} (see also 
\cite{pb96},\cite{woh95},\cite{em91} and \cite{bwf89} with 
similar results).
The relative difference is less than 0.08\% (see also Table 1).
To conclude this part, we emphasize that our results yield
a high  precision 
(relative differences of the order of 2\% and less) in the whole
$U$ range, which has not been achieved by any of the other methods
discussed. 

Let us emphasize that the presented method is not just a clever
way of making finite size extrapolations. To show that, we consider
the two-dimensional antiferromagnetic Ising model $H=\sum_{ij}
S_i^zS_j^z$ on a square lattice with spin 1/2. The ground state
energy per site is given by -1/2, which is simply the result of
each site having two bonds, each bond contributing with an energy
of $-1/4$. Any finite size calculation of this energy
would deviate from the exact value, because it would involve the
energy of spins at the boundary of the finite size cluster, where
the number of contributing bonds per site is less than 2. 
However it is an easy task to check, that our method gives
$E(C_0)=-1/4$, $I(C_1)=-1/4$, $I(C_n)=0$ for $n\neq0,1$. Thus
our expansion terminates after the first order, and in this order we
obtain precisely $\epsilon = -1/2$. 
This should make clear
that incremental expansions are at any stage yielding results for
the infinite lattice.

Using exact diagonalizations we could extend the calculations 
even further up to 6th order, i.e. up to adding six missing bonds.
This needs a careful classification of all contributing configurations
and their weight factors. We are currently working on this project.
Notice the extremely high precision which we achieve already
in 3d order, where the largest systems we have to deal with consist
of 8 sites. Without any extrapolation we obtain a precision which
e.g. in QMC is achieved by considering systems with size up to
$16 \times16 = 256$ sites and additional extrapolations
(cf. Fig.5 in \cite{wsslgs89}). The reason for this
is the fact that we use a scheme which at each level describes
an infinite system, and accounts for the important topological structures
through the weight factors. This appears to be much better than
just to consider a finite lattice with a certain size.
Having the ground state energy with that accuracy, we plan next
to account for the dimerization of a two-dimensional Hubbard-Peierls
system. Work is in progress.
\\
\\
\\
We thank P. Fulde for continuous support and
K. W. Becker, S. Denisov, P. Fulde and R. Hetzel  
for helpful discussions.

\newpage
\noindent
TABLE 1.
\\
Comparison of $\epsilon$ from 3rd order increments with 
QMC results and exact value for $U=0$.
\vspace*{2cm}
\\
\begin{tabular}{|c|c|c|}
\hline
$U$ & $\epsilon$ (ref) & $\epsilon$ (3rd order inc.) \\
\hline
\hline
0 & -1.621 (exact) & -1.634 \\ 
\hline
1 &  -1.376 (\cite{msswb90}) & -1.400 \\
\hline
2 &  -1.172 (\cite{msswb90}) & -1.191 \\
\hline
4 &  -0.841 (\cite{msswb90}) & -0.856 \\
\hline
$\infty$ (Heisenberg) &  -0.669 (\cite{kjr92})
 & -0.670 \\
\hline
\end{tabular}

\newpage
\noindent
FIGURE CAPTIONS
\\
\\
\\
Fig.1
\\
Schematic representation of $H_0$. Circles - lattice sites,
lines - bonds kept from $H$.
\\
\\
\\
Fig.2
\\
Different configurations (see text).
\\
\\
\\
Fig.3
\\
Same as Fig.2
\\
\\
\\
Fig.4
\\
Same as Fig.2
\\
\\
\\
Fig.5
\\
$\epsilon$ versus $U$. Solid line - result from third incremental
order; filled circle - exact value at $U=0$; open squares -
QMC results \cite{msswb90}; filled diamonds - projection operator
results \cite{pb96}.
\\
Inset: Dependence of incremental contributions to $\epsilon$ on
$U$. Solid line - 0th order; thick dashed line - 1st order;
thick long dashed line - 2nd order; dotted line - 3rd order.


\newpage

\begin{thebibliography}{10}

\bibitem{pf95}
P.~Fulde.
\newblock {\em Electron Correlations in Molecules and Solids. 
Third enlarged edition}.
\newblock Springer Berlin, 1995.

\bibitem{kf98}
K.~Kladko and P.~Fulde.
\newblock {\em Int. J. Quantum Chem.}, 66:377, 1998.

\bibitem{mkf98}
J.~Malek, K.~Kladko, and S.~Flach.
\newblock {\em JETP Letters}, 67:1052, 1998 
({\em Pis'ma Zh. Eksp. Teor. Fiz.}, 67:994, 1998).

\bibitem{msswb90}
A.~Moreo, D.~J. Scalapino, R.~L. Sugar, S.~R. White, and N.~E. Bickers.
\newblock {\em Phys. Rev. B}, 41:2313, 1990.

\bibitem{pb96}
G.~Polatsek and K.~W. Becker.
\newblock {\em Phys. Rev. B}, 54:1637, 1996.

\bibitem{kjr92}
K.~J. Runge.
\newblock {\em Phys. Rev. B}, 45:7229, 1992.

\bibitem{woh95}
Z.~Weihong, J.~Oitmaa, and C.~J. Hammer.
\newblock {\em Phys. Rev. B}, 52:10278, 1995.

\bibitem{em91}
E.~Manousakis.
\newblock {\em Rev. Mod. Phys.}, 63:1, 1991.

\bibitem{bwf89}
K.~W. Becker, H.~Won, and P.~Fulde.
\newblock {\em Z. Phys. B}, 75:335, 1989.

\bibitem{wsslgs89}
S.~R. White, D.~J. Scalapino, R.~L. Sugar, E.~Y. Loh, J.~E. Gubernatis, and
  R.~T. Scalettar.
\newblock {\em Phys. Rev. B}, 40:506, 1989.

\end{thebibliography}
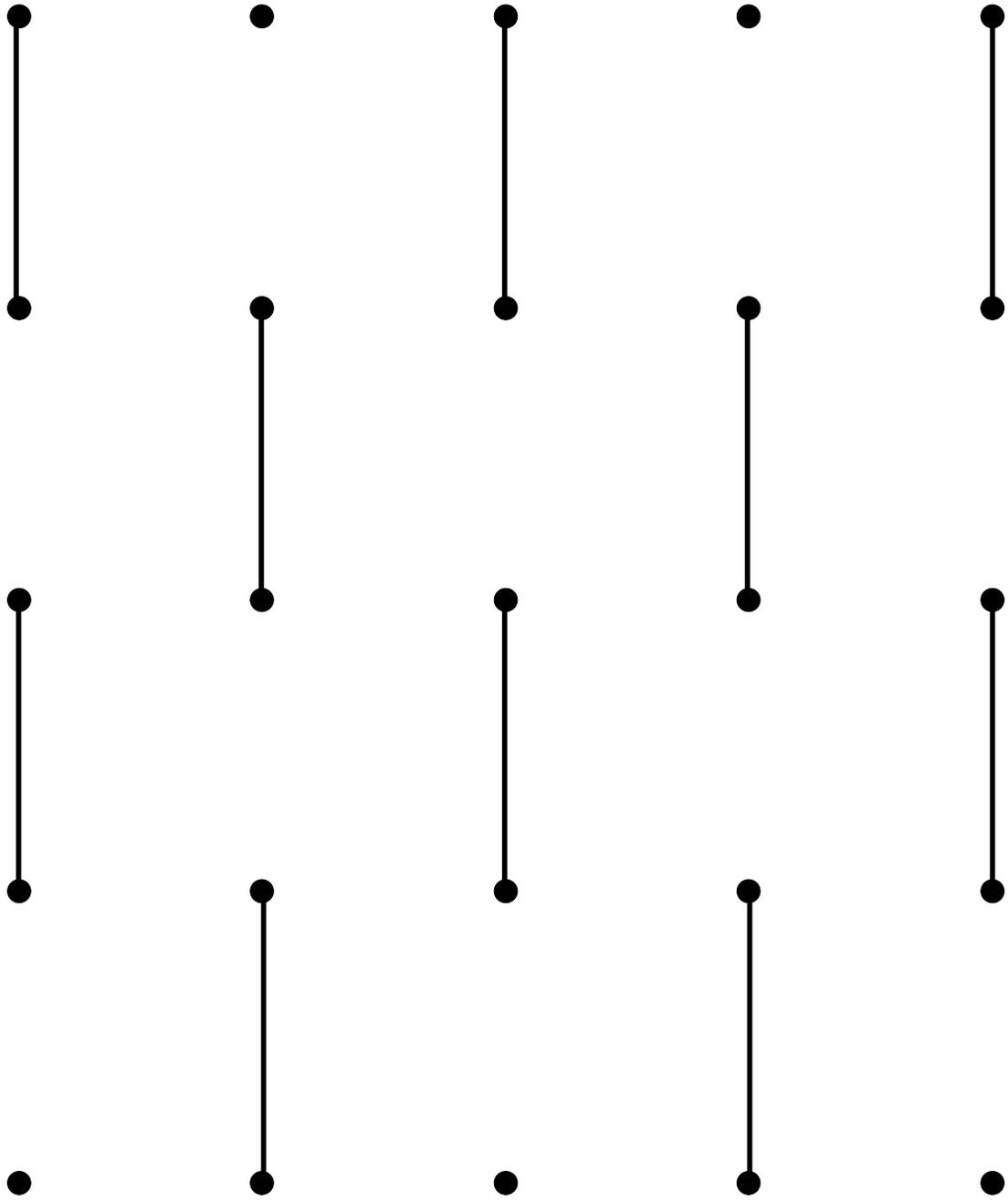
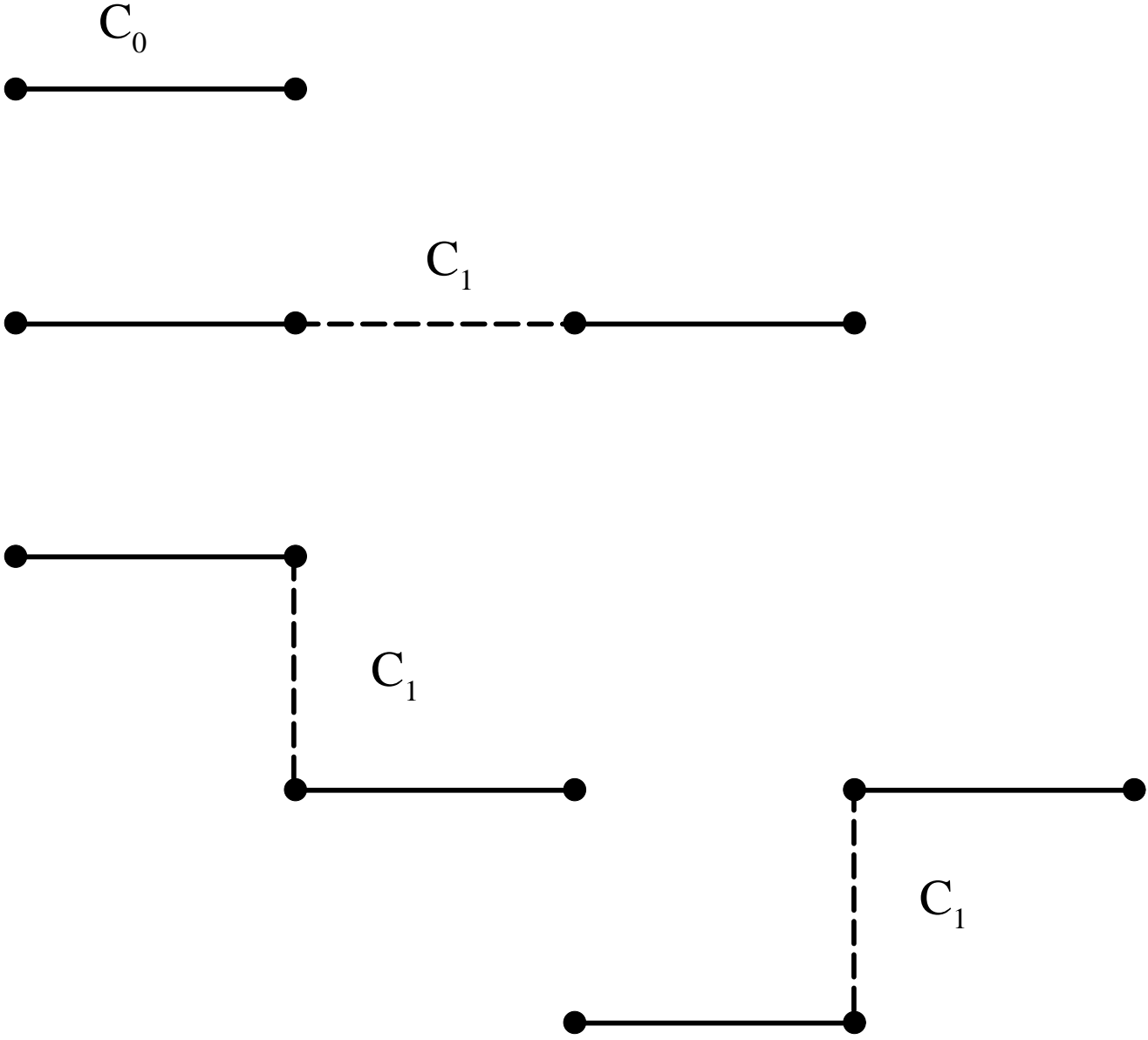
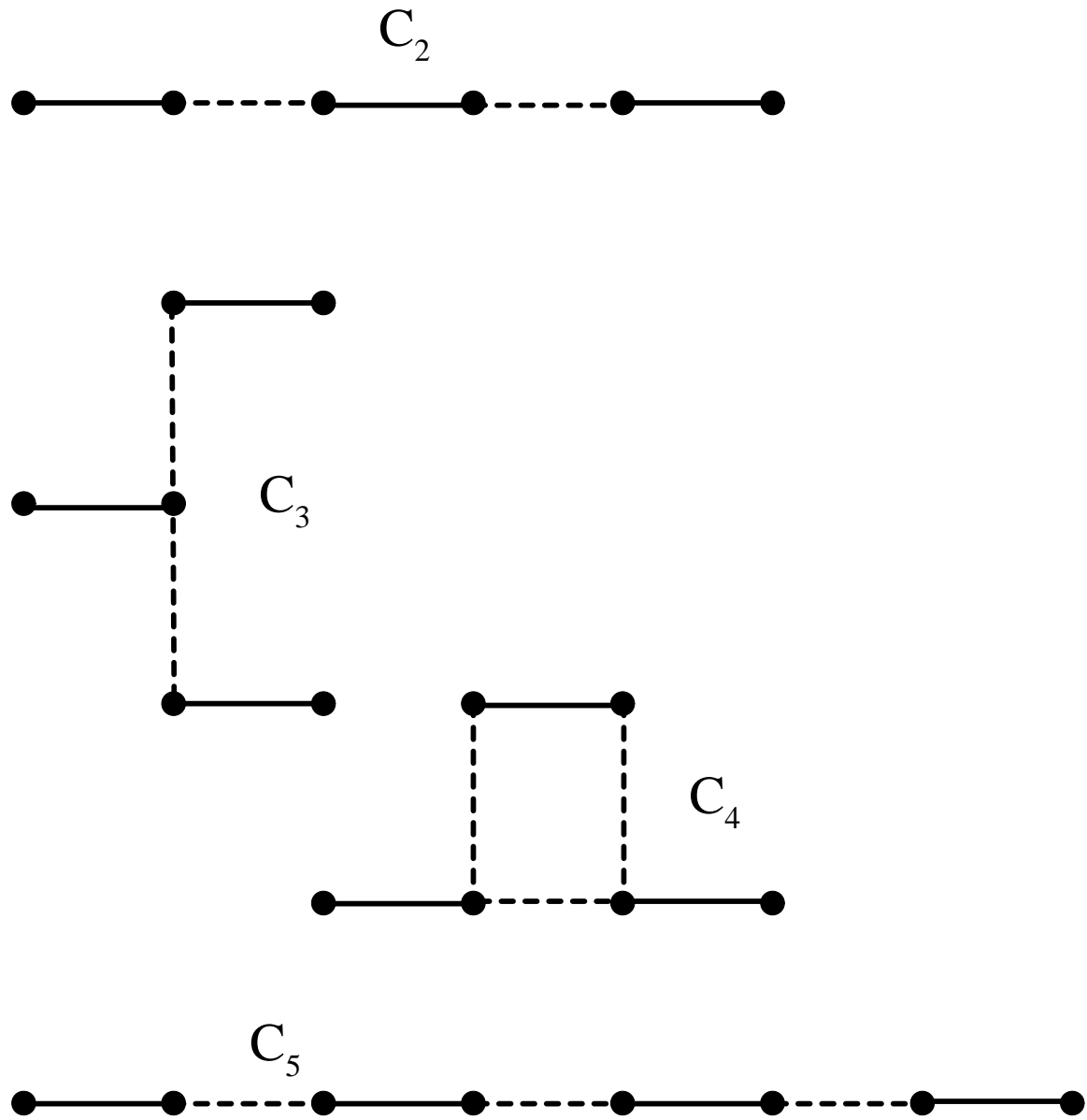
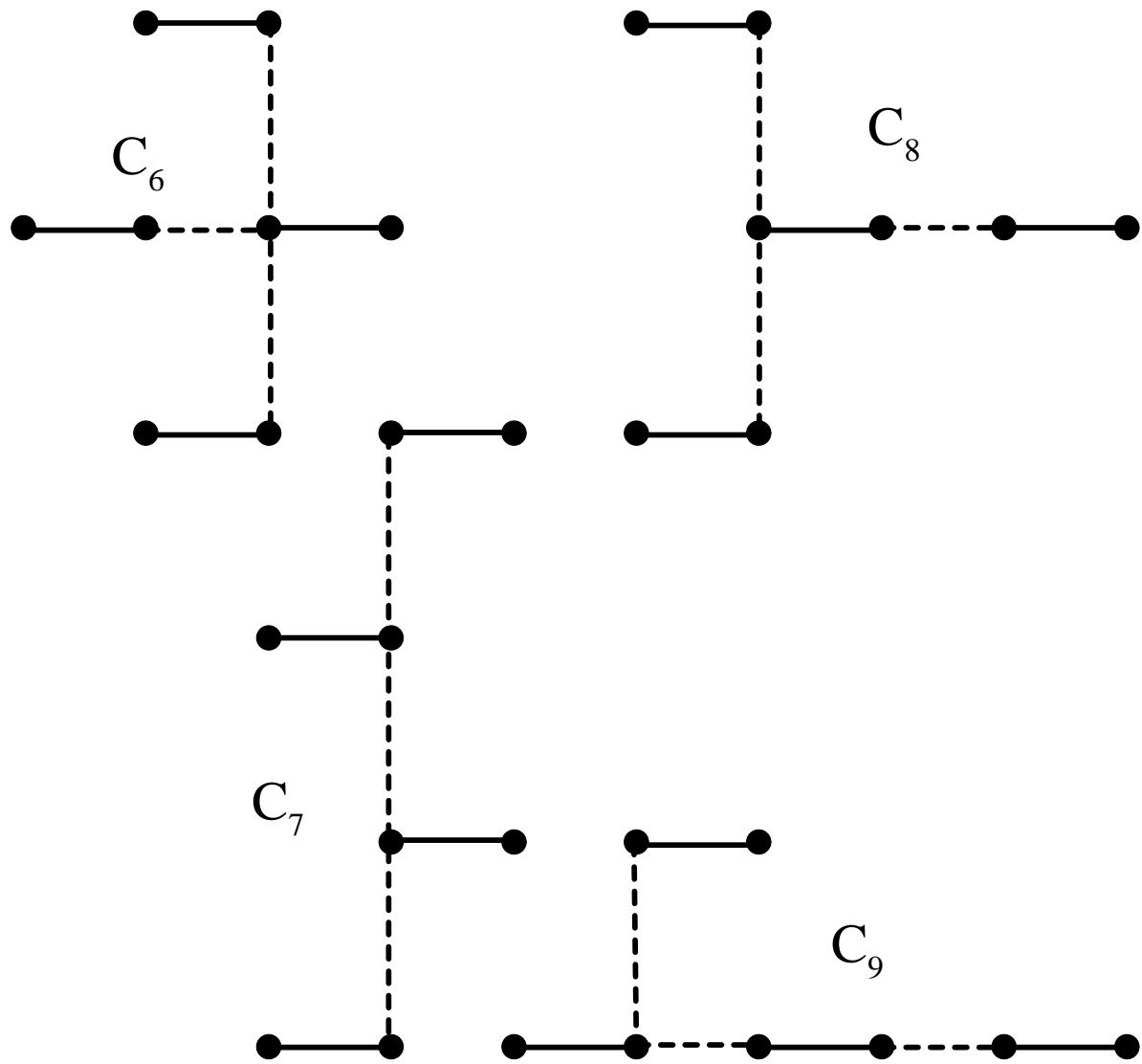
\end{document}